\documentclass[prb,preprint, amsmath,amssymb]{revtex4-1} 

\usepackage{booktabs}
\usepackage{graphicx}
\usepackage{subfigure}
\usepackage[section]{placeins}
\usepackage{soul}
\usepackage{orcidlink}
\usepackage{hyperref}   
\hypersetup{
	colorlinks=true, 
	linkcolor=blue, 
	citecolor=blue, 
	urlcolor=magenta, 
	pdfborder={0 0 0} 
}

\begin{document}
\title{Induced electromotive force of a thin metal rod in the alternating electromagnetic field of Helmholtz coil: experimental results and theoretical analysis}

\author{Yilin Shao}
\thanks{These authors contributed equally to this work.}
\affiliation{
	School of Science, Huzhou Normal University, Huzhou 313000, Zhejiang , China}
\author{Minghan Gao}
\thanks{These authors contributed equally to this work.}
\affiliation{
	School of Science, Huzhou Normal University, Huzhou 313000, Zhejiang , China}
\author{Baiqing Li}
\affiliation{
	School of Science, Huzhou Normal University, Huzhou 313000, Zhejiang , China}
\author{Xiaoguang Li}
 \affiliation{   School of Intelligent Manufacturing, Huzhou University, Huzhou 313000, Zhejiang , China}
 
\author{Chengfu Mu}
\email{muchengfu@huznu.edu.cn}
\thanks{Author to whom correspondence should be addressed.}

\affiliation{
	School of Science, Huzhou Normal University, Huzhou 313000, Zhejiang , China}

\begin{abstract}
	 We apply a thin metal rod (copper rod) as a probe in the alternating magnetic field generated by a Helmholtz coil, and measure the variation of the induced electromotive force (EMF) on the metal rod at different radial positions of Helmholtz coil. Experimental results show that the induced EMF is zero when the center of the metal rod passes through the center of the cylindrical magnetic field inside the Helmholtz coil. When the metal rod is displaced from the center of field to different radial positions, the induced EMF gradually increases from zero, reaches a maximum at a certain position, and then decreases monotonically as the radial distance continues to increase. At the position where the induced EMF reaches its maximum, the metal rod intersects the radial cross-section of the internal magnetic field of the Helmholtz coil at two points, with a small central portion of the rod located inside the Helmholtz coil and the two end portions outside the coil. To explain the experimental phenomena, we construct four simplified models of the magnetic field distribution based on the actual field distribution of the Helmholtz coil to quantitatively investigate the radial variation of the induced EMF along the metal rod. Our theoretical results show that the four models yield similar results and can all qualitatively explain the experimental data curves, particularly reproducing well the variation trend of the induced EMF and the position of the extremum point. The result of piecewise function fitting model is quantitatively in good agreement with the experimental data. This work is also very much helpful and instructive for undergraduate-level students.
\end{abstract}

\maketitle

\section {Introduction}
\label{sec1}

All kinds of Helmholtz coil devices are widely used in many frontier fields of modern scientific research, such as the calibration of various magnetic field meters and sensors \cite{Garczarek2024,DEMELO20091330,Ratajczak_2020}, the medical therapy for real-time circulating tumor cell separation\cite{KANG2024115229}, particle detector\cite{Bestmann2017}, and the study of diamagnetism in laser-produced plasma \cite{Behera2019}. There are many experimental instruments capable of generating homogeneous magnetic fields. For example, a uniformly and closely wound long solenoid produces a highly spatially uniform magnetic field inside the solenoid. However, a closely wound solenoid lacks openness, making it inconvenient to place experimental devices inside the magnetic field for measurement operations. Unlike the closely wound solenoid, the traditional Helmholtz coil generates a uniform magnetic field in the region near its axis through its unique double-coil structure. Its openness gives it unique advantages in engineering measurements and experimental instrument operations, making it very suitable for research work requiring a uniform magnetic field. In recent years, various new types of Helmholtz coil devices have been developed to improve the range and stability of the uniform magnetic field of traditional Helmholtz coils, such as three-axis Helmholtz coils \cite{zhang2023}, and new Helmholtz coil systems composed of four square coils \cite{ZHU2025101944}, greatly expanding their application range \cite{Li2024,Hurtado2016,Thrysoe_2025,Chen2022,Zhu2023}.

In the field of precision measurement, in addition to the direct use of magnetic fields of various intensities as direct detection signals, the induced EMF generated by electromagnetic induction serves as a key measurement signal and has been widely applied in engineering measurement and experimental instrument design. Various detection coils, probes as typical application carriers of the electromagnetic induction principle, have become indispensable key components. They utilize the induced EMF generated by electromagnetic induction as the detection signal, and then efficiently convert weak electromagnetic signals into electrical signals through signal amplification methods, thereby achieving precision measurement \cite{Behera2019,Datta2022}. The precise measurement and investigation of the variation law of the induced EMF of detection coils or probes in magnetic fields have important theoretical and practical application for the measurement of various experimental signals.

In this paper, we construct a cylindrical spatial magnetic field using a Helmholtz coil and place a thin metal rod as a probe inside and outside the coil. An induced EMF is generated on the metal rod. We place the metal rod in different radial regions of the cylindrical magnetic field, measure the variation of the induced EMF on the rod with radial distance, and determine the position of the metal rod in the magnetic field when the induced EMF reaches its maximum value. We establish four simplified theoretical models of the magnetic field based on the theoretical simulation of Helmholtz coil to explain the variation law of the induced EMF on the thin metal rod observed experimentally. The experimental and theoretical research in this paper helps to expand the application range of Helmholtz coils, making their application not limited to the usual central uniform magnetic field region, and provides an extensive basis for the engineering measurements and calibration of magnetic field meters, the design of magnetic field sensors, and the optimization of other experimental instruments.

 \section{ Experimental setup} 
 \label{Sec2}
 
\begin{figure}[htbp]
   \subfigure[]{ 
    \centering
         \begin{minipage}[t]{0.38\textwidth}
         \vspace{0pt}  
        \centering
        \includegraphics[width=\linewidth]{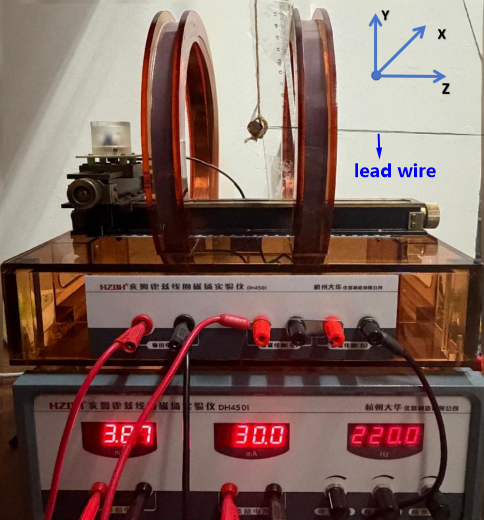}
        \label{fig:setup1}
        \\[4pt]
        \end{minipage}}
    \subfigure[]{ 
    \centering
    \begin{minipage}[t]{0.38\textwidth}
        \vspace{0pt}  
        \centering
        \scalebox{1}[1.15]{\includegraphics[width=\linewidth]{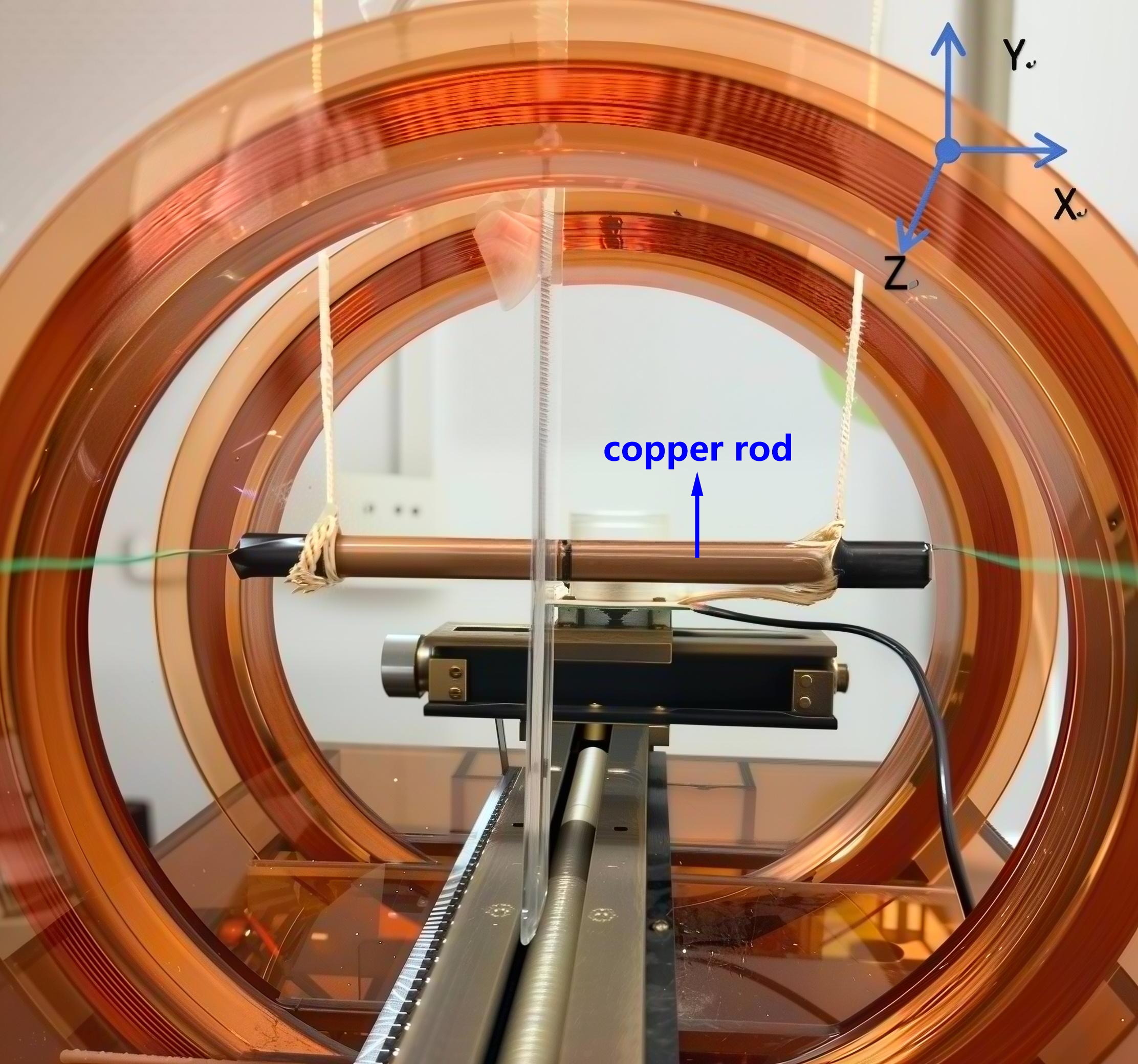}}
        \label{fig:setup2}
        \\[4pt]
        \end{minipage}
       }
    \caption{ Helmholtz coil and thin metal rod setup: (a) side view. (b) front view.}
    \label{fig:setup}
\end{figure}

 The Helmholtz coil used in this experiment consists of two parts: the excitation coil assembly and the magnetic field measuring instrument. The excitation coil assembly comprises two excitation coils, each with an effective radius $R = 10.5\ \mathrm{cm}$ (inner radius $10.0\ \mathrm{cm}$, outer radius $11.0\ \mathrm{cm}$), $400$ turns per coil, and a center-to-center separation of $10.5\ \mathrm{cm}$. The excitation alternating current (AC) applied to the Helmholtz coil has a frequency of $220\ \mathrm{Hz}$ and an effective root mean square (RMS) value of $30\ \mathrm{mA}$. Due to the good electrical conductivity of the copper rod, we choose it to measure the EMF in this experiment. The thin copper rod used has a length of $2l = 15\ \mathrm{cm}$ ($l < R$) and a diameter of $1\ \mathrm{cm}$ (negligible compared to the coil radius $R$). The experimental setup is shown in figures.~\ref{fig:setup1} and \ref{fig:setup2}.

We use a high-precision multimeter with an accuracy of 0.001 mV in the AC millivolt voltage range to measure the induced EMF on the thin metal rod. Since the direction of magnetic field inside the Helmholtz coil is always along its axis ($z$ diretion), in order to avoid interference caused by the additional EMF generated by the straight lead wires connected to both ends of the metal rod, we stretch the lead wires straight and parallel to the magnetic field direction. In this way, the straight wires do not cut the magnetic lines of force, thus reducing the influence of the induced EMF from the lead wires on the measurement for the thin metal rod, as shown in figure~\ref{fig:setup}. To mitigate the position shift of the metal rod induced by tension on the lead wires, we employ a ruler as a baffle to ensure the metal rod moves strictly up and down along the baffle in the $xoy$ plane. During the experiment, we try our best to keep the metal rod parallel to the $x$-axis, with its center always lying on the $y$-axis. We also minimize measurement errors arising from the tilt or deflection of the metal rod.

At the beginning, the metal rod lies in the $xoy$ plane ($z=0$), its center coincides with the center of the Helmholtz coil ($y=0$), and it keeps parallel to the $x$ axis throughout the subsequent experiment. We measure how the EMF value of the stationary metal rod varies at different $y$ positions (note that this EMF originates from the time-varying excitation current). The acquired experimental data are listed in Table~\ref{tab:data}.

\begin{table}[htbp]
\centering
\small
\setlength{\tabcolsep}{3.5pt}
\caption{Experimental data of position $y$ and induced EMF $\varepsilon$.}
\label{tab:data}
\begin{tabular}{ccccccccccc}
\hline
Position No. & 1 & 2 & 3 & 4 & 5 & 6 & 7 & 8 & 9 & 10 \\
\hline
$y$ (cm) & 0.0 & 1.0 & 2.0 & 2.7 & 3.6 & 4.2 & 4.8 & 5.5 & 6.3 & 6.9 \\
$\varepsilon_{AB}$ (mV) & 0.000 & 0.099 & 0.195 & 0.295 & 0.396 & 0.492 & 0.526 & 0.593 & 0.657 & 0.694 \\
\hline
Position No. & 11 & 12 & 13 & 14 & 15 & 16 & 17 & 18 & 19 & 20 \\
\hline
$y$ (cm) & 7.7 & 8.5 & 9.3 & 9.8 & 10.6 & 11.5 & 12.4 & 13.2 & 14.4 & 15.1 \\
$\varepsilon_{AB}$ (mV) & 0.750 & 0.780 & 0.805 & 0.807 & 0.793 & 0.790 & 0.752 & 0.728 & 0.658 & 0.628 \\
\hline
Position No. & 21 & 22 & 23 & 24 & 25 & 26 & 27 & 28 & 29 & 30 \\
\hline
$y$ (cm) & 15.3 & 17.0 & 17.5 & 18.0 & 18.4 & 20.1 & 21.4 & 23.3 & 24.0 & 25.0 \\
$\varepsilon_{AB}$ (mV) & 0.605 & 0.522 & 0.480 & 0.460 & 0.455 & 0.400 & 0.360 & 0.310 & 0.300 & 0.285 \\
\hline
\end{tabular}
\end{table}

\begin{figure}
    \includegraphics[width=0.9\linewidth]{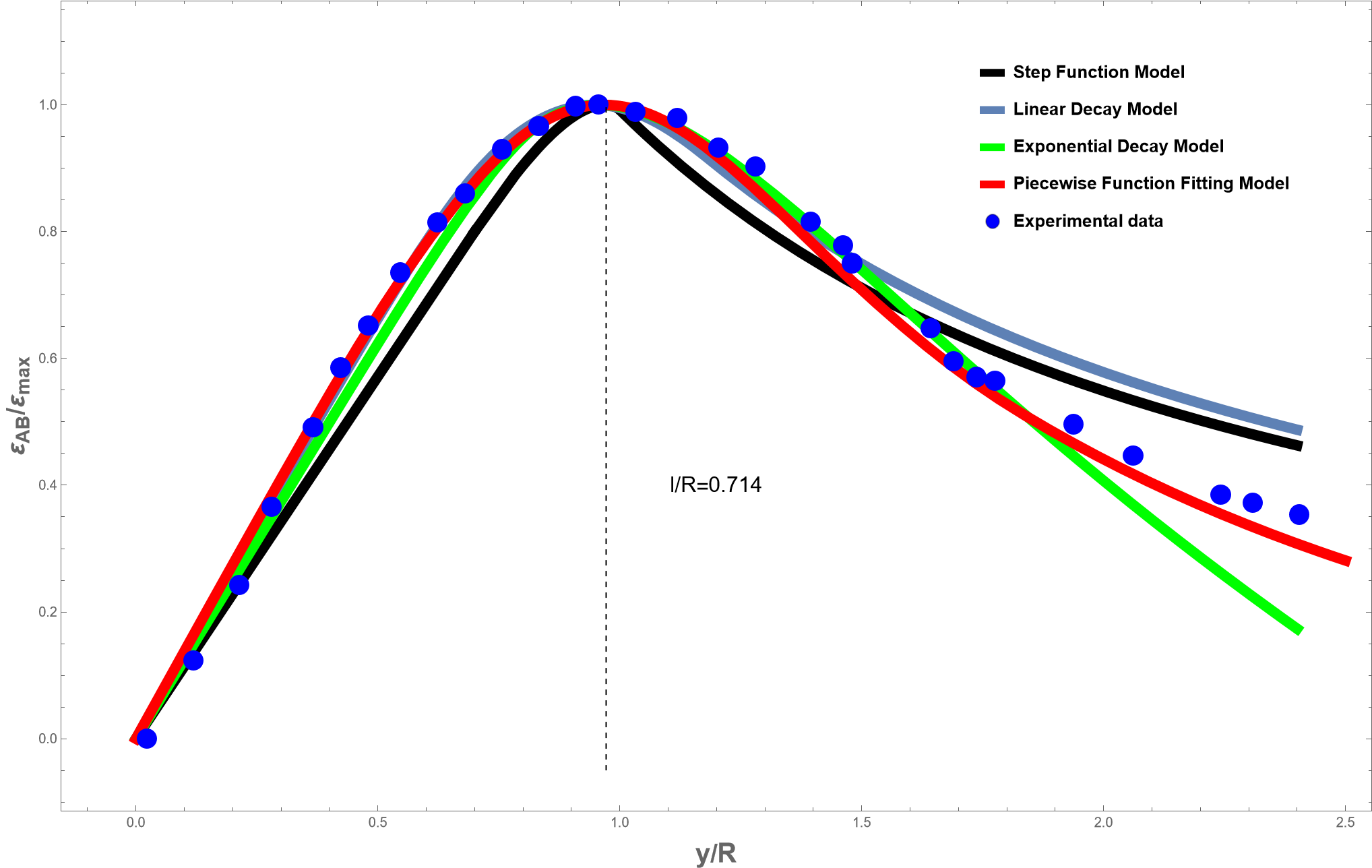}
    \caption{Variation of the EMF in the middle plane of the Helmholtz coil with radial distance $y$ (dashed line indicates the position of the maximum value of $\varepsilon_{AB}$).}
    \label{fig:emf}
\end{figure}

In order to determine the effect of the additional EMF that may be generated by the straight lead wires deviating from the $z$ axis on the metal rod EMF, we intentionally let the straight lead wires deviate from the $z$ axis of the Helmholtz coil and found that the error caused by the deviation of the straight lead wires from the axis direction is generally small, less than $10\%$, and does not affect our discussion. In order to compare with the theoretical results later, we use dimensionless physical quantities for plotting. For the vertical coordinate $y$, we normalize it using the average radius $R$ of the Helmholtz coil. For the measured EMF, we normalize it using the maximum measured EMF. The experimental results are shown as blue dots in figure~\ref{fig:emf}.

From figure.~\ref{fig:emf}, it can be seen that the induced EMF on the metal rod is zero at the center of the Helmholtz coil. As the metal rod moves gradually along the $y$ direction, its induced EMF increases accordingly. When $y=9.8$~cm, i.e., $y/R=0.933$, the EMF reaches its maximum value of $0.807$~mV, which corresponds to the position where the normalized vertical coordinate equals $1$ in figure~\ref{fig:emf}. As $y$ further increases up to approximately $2.4$ times the coil radius, the induced EMF decreases monotonically. In addition, we changed the length of the straight lead wires during the experiment, with all wire ends placed outside the axial range of the Helmholtz coil. The experimental results only changed a little, indicating that the measured EMF primarily originates from the metal rod induced by the alternating excitation current, while the additional EMF from the lead wires has a minor influence on the experimental results. To interpret the observed phenomena, four simplified radial magnetic field models of the Helmholtz coil are adopted to calculate the induced EMF on the metal rod. The theoretical results are plotted as four solid lines in figure~\ref{fig:emf}, which are discussed in detail in the following sections.

\section{Experimental data of the magnetic field of Helmholtz coil and their theoretical models}
\label{Sec3}

\subsection{Experimental data and theoretical results of the radial distribution of the magnetic field}
\begin{figure}
    \centering
    \includegraphics[width=0.9\linewidth]{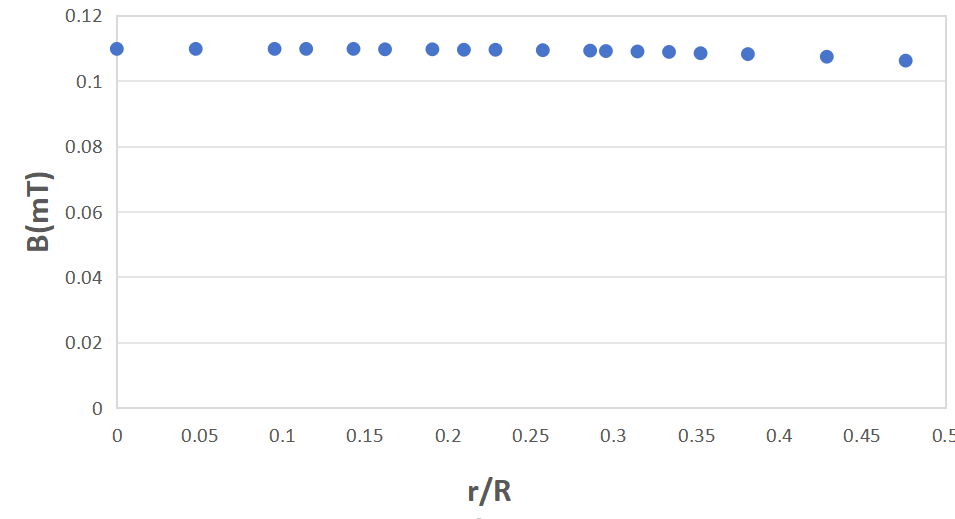}
    \caption{Experimentally measured radial magnetic field distribution of the Helmholtz coil.}
    \label{fig:magnetic}
\end{figure}

\begin{figure}
   \subfigure[]{ 
    \centering
         \begin{minipage}[b]{0.42\textwidth}
         \vspace{0pt}  
        \centering
        \includegraphics[width=\linewidth]{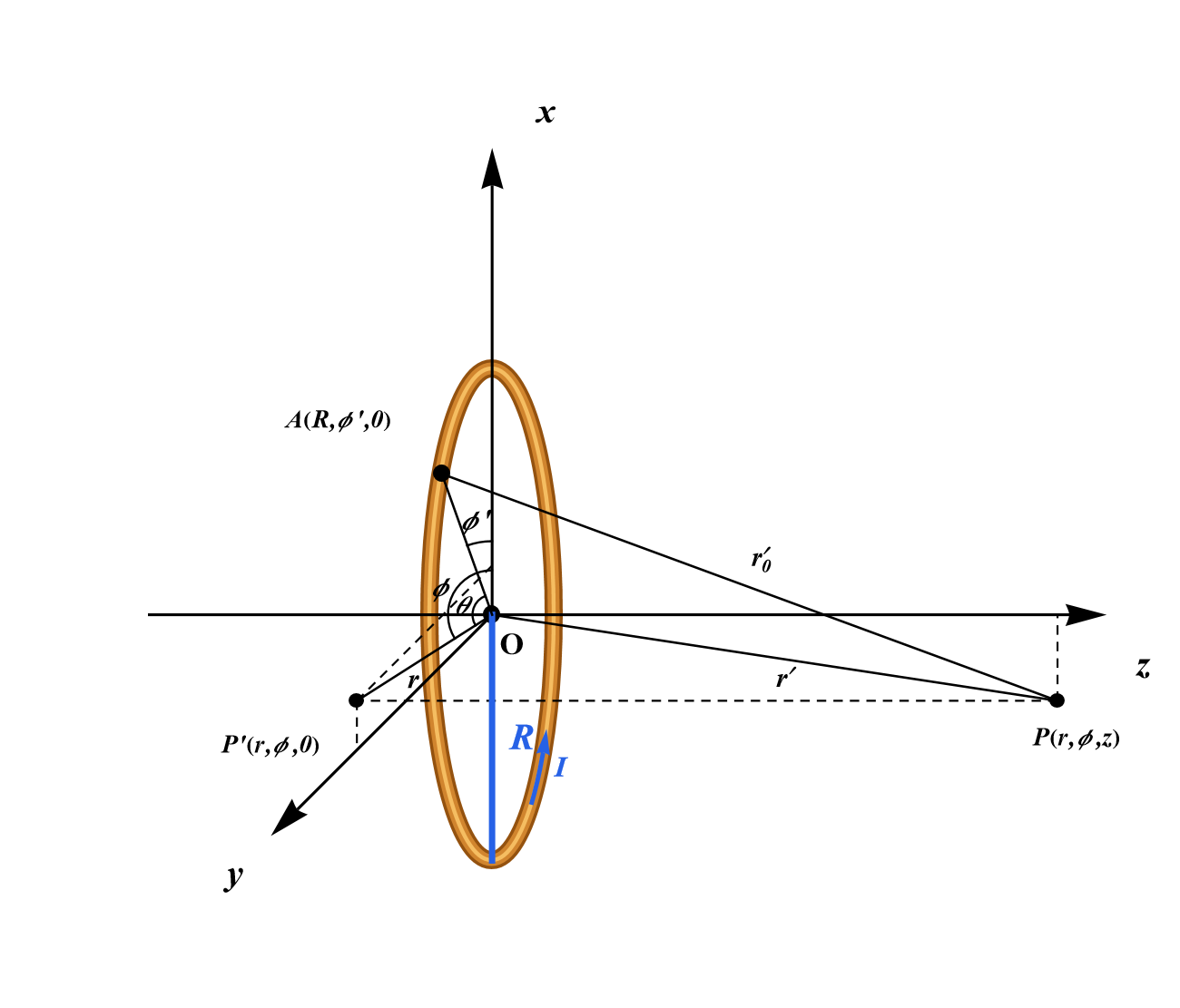}
        \label{fig:4a}
        \\[4pt]
        \end{minipage}}
    \subfigure[]{ 
    \centering
    \begin{minipage}[b]{0.42\textwidth}
        \vspace{0pt}  
        \centering
        \scalebox{1}[1.0]{\includegraphics[width=\linewidth]{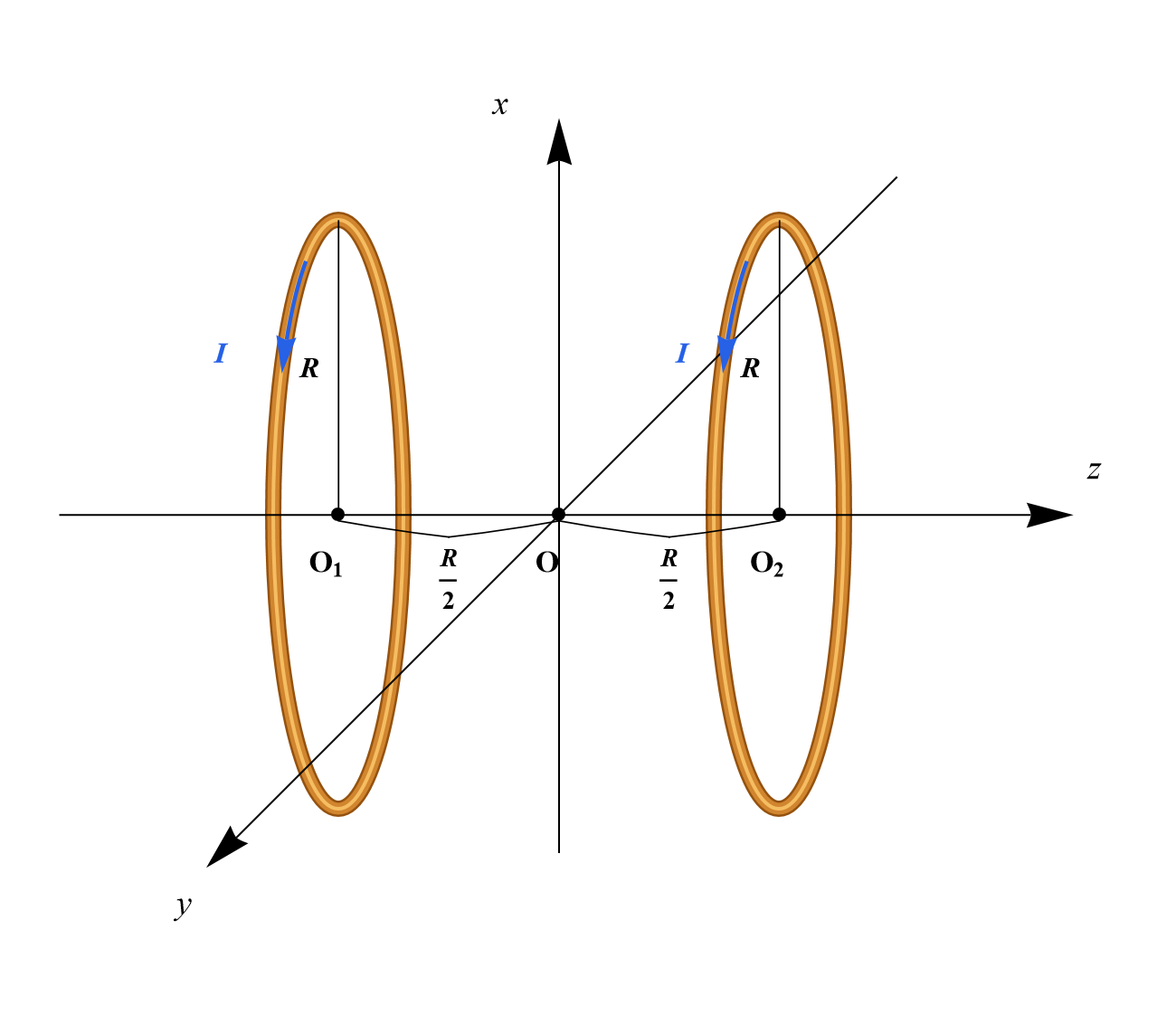}}
        \label{fig:4b}
        \\[4pt]
        \end{minipage}
       }
    \caption{(a) Magnetic field generated by a circular current at an arbitrary spatial point; (b) Schematic diagram of the Helmholtz coil.}
    \label{fig:4}
\end{figure}

First, we measure the radial magnetic field distribution of of the Helmholtz coil. Due to the limited range of the radial movement knob of the experimental instrument, we can only measure the distribution within a radial distance of $5$~cm (close to half the radius $R$ of the Helmholtz coil, with a radial distance resolution of $1$~mm). Beyond this range, the magnetic field distribution is replaced by the theoretical results of the Helmholtz coil model. We measure the radial distribution point by point. The experimental results are shown in Table~\ref{tab:magnetic}. Here, $U$ is the effective voltage measured by the detection coil in the Helmholtz coil, from which the magnetic field value at each point can be calculated. For intuitive discussion, we also present these experimental data in figure ~\ref{fig:magnetic}. From figure ~\ref{fig:magnetic}, we can see that the actual magnetic field strength $B$ decreases very little until it approaches half the magnetic field region $R/2$, less than $4\%$. The experiment shows that within our measurement range, the magnetic field of the Helmholtz coil has good spatial uniformity and can be regarded as a uniform magnetic field, which is also the main application region of the Helmholtz coil.
\begin{table}[htbp]
\centering
\small
\setlength{\tabcolsep}{3.5pt}
\caption{Experimental data of magnetic field distribution along radial direction.}
\label{tab:magnetic}
\begin{tabular}{cccccccccc}
\hline
\textbf{Position No.} & 1 & 2 & 3 & 4 & 5 & 6 & 7 & 8 & 9 \\
\hline
$r$ (mm) & 0 & 5 & 10 & 12 & 15 & 17 & 20 & 22 & 24 \\
$U$ (mv) & 8.25 & 8.25 & 8.25 & 8.25 & 8.25 & 8.24 & 8.24 & 8.23 & 8.23 \\
$B$ (mT) & 0.1097 & 0.1097 & 0.1097 & 0.1097 & 0.1097 & 0.1096 & 0.1096 & 0.1095 & 0.1095 \\
\hline
\textbf{Position No.} & 10 & 11 & 12 & 13 & 14 & 15 & 16 & 17 & 18 \\
\hline
$r$ (mm) & 27 & 30 & 31 & 33 & 35 & 37 & 40 & 45 & 50 \\
$U$ (mv) & 8.22 & 8.21 & 8.20 & 8.19 & 8.18 & 8.15 & 8.13 & 8.07 & 7.98 \\
$B$ (mT) & 0.1093 & 0.1092 & 0.1091 & 0.1089 & 0.1088 & 0.1084 & 0.1081 & 0.1073 & 0.1061 \\
\hline
\end{tabular}
\end{table}
Since the excitation current is sinusoidal alternating current, the magnetic field generated by the Helmholtz coil varies with space and time as follows:
\begin{equation}
\vec{B}_I(\vec{r},z,t) = \vec{B}(\vec{r},z) B_t(t) = \vec{B}(\vec{r},z) \sin \omega t
\label{eq:1}
\end{equation}
where $r$ denotes the radial coordinate of the Helmholtz-coil system, and $\sin(\omega t)$ represents the time-dependent component of the magnetic field. Since the magnetic field at every point changes synchronously with time, we concentrate on the spatial dependence of the magnetic field in the following discussion. Based on the Biot--Savart law, we derive the spatial magnetic-field distribution for a single circular coil and for the Helmholtz coil; the schematic diagram is presented in figure~\ref{fig:4}. The Helmholtz coil is composed of two coaxial circular coils of identical radius $R$ carrying current $I$ in the same direction. The axial separation between the two coils equals the coil radius $R$. Taking the geometric center of the two coils as the coordinate origin, the right-hand coil is positioned at $z=+R/2$, and the left-hand coil at $z=-R/2$. The magnetic field possesses rotational symmetry, so the azimuthal component vanishes $B_\varphi=0$, leaving only the axial component $B_z$ and the radial component $B_r$. The magnetic field of Helmholtz coil at any spatial point is the vector superposition of the fields from the two circular currents.

\begin{figure}
    \centering
    \includegraphics[width=0.8\linewidth]{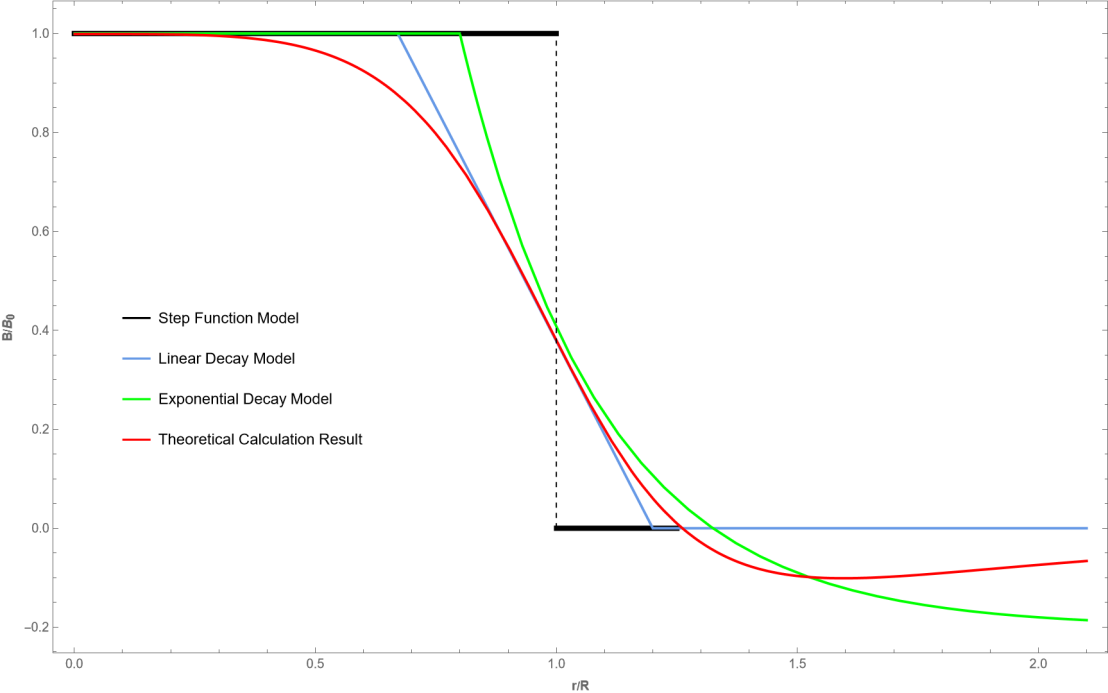}
    \caption{Comparison of normalized magnetic-field radial distribution on the mid-plane of the Helmholtz coil (the dashed line $r = R$ marks the coil-radius position).}
    \label{fig:5}
\end{figure}

\begin{figure}
    \centering
    \includegraphics[width=0.7\columnwidth]{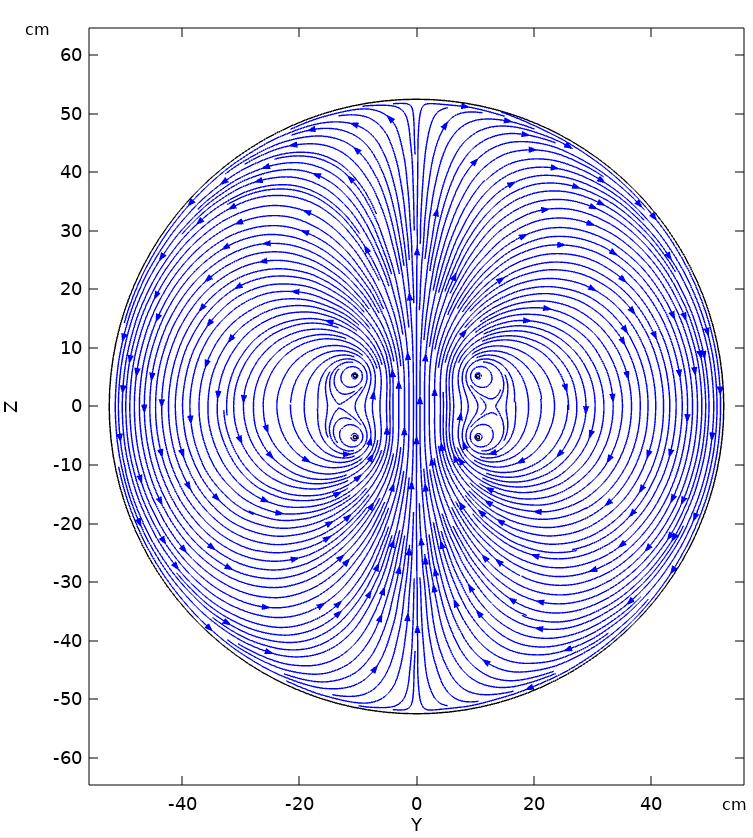}
    \caption{Full-space magnetic field distribution map.}
    \label{fig:6}
\end{figure}

Within the Helmholtz coil, the total magnetic field is predominantly aligned along the $z$ axis over a large region near the central axis. The radial $r$-component becomes non-negligible only near the coil edges and exhibits axial symmetry. Therefore, we neglect its contribution to the induced EMF on the metal rod throughout this work. Only the $z$ component of the magnetic field is regarded as the major source of the induced EMF. Unless otherwise specified, all references to the magnetic field in the following text for $B$ refer to its $B_z$ component.
As shown in figure~\ref{fig:4a}, the integral expression for the $z$ component of the magnetic field produced by a single current-carrying circular coil can be readily derived, and the result is given below \cite{Garcia2020,Stahel2009,Smythe1989}:
\begin{equation}
B_z(r,z) = \frac{\mu_0 I}{4\pi} \int_0^{2\pi} \frac{R^2 - Rr\cos\theta}{(r^2 + R^2 + z^2 - 2rR\cos\theta)^{3/2}} d\theta
\label{eq:2}
\end{equation}
where $r$ denotes the radial coordinate of the field point, and $z$ is the axial coordinate of the field point relative to the center of single coil; $\mu_0$ is the vacuum permeability, and the integration variable $\theta$ stands for the azimuthal angle. For the Helmholtz-coil configuration in figure~\ref{fig:4b}, we define $z_1 = z + R/2$ (axial distance from the field point to the left coil) and $z_2 = R/2 - z$ (axial distance from the field point to the right coil). Superposing the contributions from both coils and using equation~\eqref{eq:2}, we obtain the integral expression for the $z$-component of the magnetic field of the Helmholtz coil:
\begin{equation}
\begin{split}
B_{z,\mathrm{Helm}}(r,z) = {}& \frac{\mu_0 I}{4\pi} \int_0^{2\pi} \frac{R^2 - Rr\cos\theta}{(r^2 + R^2 + z_1^2 - 2rR\cos\theta)^{3/2}} d\theta \\
&+ \frac{\mu_0 I}{4\pi} \int_0^{2\pi} \frac{R^2 - Rr\cos\theta}{(r^2 + R^2 + z_2^2 - 2rR\cos\theta)^{3/2}} d\theta
\end{split}
\label{eq:3}
\end{equation}
Based on equation~\eqref{eq:3}, we can plot how the magnetic field magnitude of the Helmholtz coil varies with radial distance $r$, as plotted by the red solid line in figure~\ref{fig:5}. For convenience in analysis, dimensionless quantities are adopted, and the magnetic field values are normalized by its maximum value such that the maximum of the normalized field equals unity. From the red curve in figure~\ref{fig:5}, one can observe that the $z$-component of the Helmholtz coil magnetic field decays obviously starting from $0.5R$ and smoothly drops to zero near $1.26R$. As the radial distance increases further, the direction of magnetic field reverses; its magnitude first rises and then decreases. Theoretically, this magnetic field vanishes at infinity.
To acquire the full-space magnetic field distribution of the Helmholtz coil, we carried out numerical simulations with COMSOL, and the results are presented in figure~\ref{fig:6}. It can be observed that the region near the central axis over a considerable extent can be approximated as a uniform magnetic field oriented along the $z$ axis. Far outside the Helmholtz coil, the magnetic field points along the negative $z$ direction.

\subsection{Theoretical models of the radial distribution magnetic field}

The theoretical expression equation~\ref{eq:3} for the magnetic field distribution of the Helmholtz coil involves elliptic functions and is very complicated, which is not convenient for us to theoretically calculate the induced EMF on the metal rod. In order to obtain a simple functional expression for the induced EMF, we establish the following four simplified models of the radial magnetic field in the mid-plane of Helmholtz coil ($z=0$) based on the magnetic field distribution (red line) in figure~\ref{fig:5}: the step function model, the linear decay model, the exponential decay model, and the piecewise function fitting model. This greatly simplifies our EMF calculation, and semi-analytical expressions can be obtained. We find that the results for the EMF on the metal rod obtained from different models are qualitatively consistent with the experimental results. We will discuss them separately in the following.
\subsubsection{Model A: step function model}

To capture the physical essence of EMF variation for the metal rod inside the Helmholtz coil, the simplest model treats the entire interior of the Helmholtz coil as a uniform magnetic field $B_0$, while the magnetic field outside the coil is set to zero. This is referred to as the step function model assumption:
\begin{equation}
\frac{B(r)}{B_0} = 
\begin{cases}
1, & r \le R \\
0, & r > R
\end{cases}
\label{eq:4}
\end{equation}
This profile is represented by the black solid line in figure~\ref{fig:5}. Strictly speaking, the step function model describes the magnetic field distribution of an infinitely long tightly wound solenoid rather than a Helmholtz coil. Nevertheless, it serves as a reasonable first-order approximation for model simplification; hence it can also be named the infinitely long tightly wound solenoid model. This model yields a compact calculation for the EMF induced on the metal rod. Its predicted results are qualitatively consistent with experimental observations, which facilitates insight into the underlying physical mechanism. Detailed analysis will be presented in the following section \ref{Sec4}.

\subsubsection{Model B: Linear decay model}

According to the numerical simulation results for the Helmholtz coil magnetic field (red solid line in figure~\ref{fig:5}), the magnetic field inside the coil is uniform only within a limited cylindrical region near the central axis. The magnetic field starts to decay near $r=0.378R$ and falls to zero around $r=1.26R$, instead of dropping abruptly to zero at a single radial position. To characterize the continuous transition from the ``uniform magnetic field region'' to the ``zero field region'' near the coil edge, we modify the step function model. For better agreement with the simulated magnetic field profile, a linear function is adopted to connect the segments at $r/R=0.6709$ and $r/R=1.2$, so that the magnetic field decays linearly, as illustrated by the blue straight line in figure~\ref{fig:5}. This is defined as the linear decay magnetic field model. This model captures the transition behaviour with a minimal number of parameters and simplifies the subsequent calculation of the induced EMF on the metal rod. The fitted magnetic field distribution is given below:
\begin{equation}
\frac{B(r)}{B_0} = 
\begin{cases}
1, & 0 \le r/R < R_{21} \\
ar + b, & R_{21} \le r/R < R_{22} \\
0, & R_{22} \le r/R
\end{cases}
\label{eq:5}
\end{equation}
where $R_{21}=0.6709$ and $R_{22}=1.2$. The coefficients $a=-1.89$ and $b=2.268$ are obtained from fitting.

\subsubsection{Model C: Exponential decay model}

Since the magnetic field decay inside the Helmholtz coil deviates from a linear behaviour, we can further modify the linear decay model. By analogy to the distribution function of fermions at finite temperature, an exponential function is adopted to connect the uniform inner magnetic field and the reversed outer magnetic field of the Helmholtz coil. A shift is introduced to the exponential decay term for better agreement with theoretical simulation results of Helmholtz coil (red curve in figure~\ref{fig:5}). The fitted profile is plotted as the green curve in figure~\ref{fig:5}. The explicit form reads:
\begin{equation}
\frac{B(r)}{B_0} = 
\begin{cases}
1, & 0 \le r/R < R_{31} \\[6pt]
c e^{\frac{R-r}{R_{00}}} + d, & r/R \ge R_{31}
\end{cases}
\label{eq:6}
\end{equation}
where $R_{31} = 0.8$. The three fitted parameters are $R_{00} = 0.293R$, $c = 0.606$ and $d = -0.2$. As illustrated by figure~\ref{fig:5}, the exponential decay model can reproduce the reversed magnetic field region outside the coil within a certain radial range. Nevertheless, for $r/R \ge 1.6$, the reversed field magnitude predicted by this model deviates significantly toward larger values compared with theoretical simulation.

\subsubsection{Model D: the piecewise function fitting model}

\begin{figure}
    \includegraphics[width=0.8\linewidth]{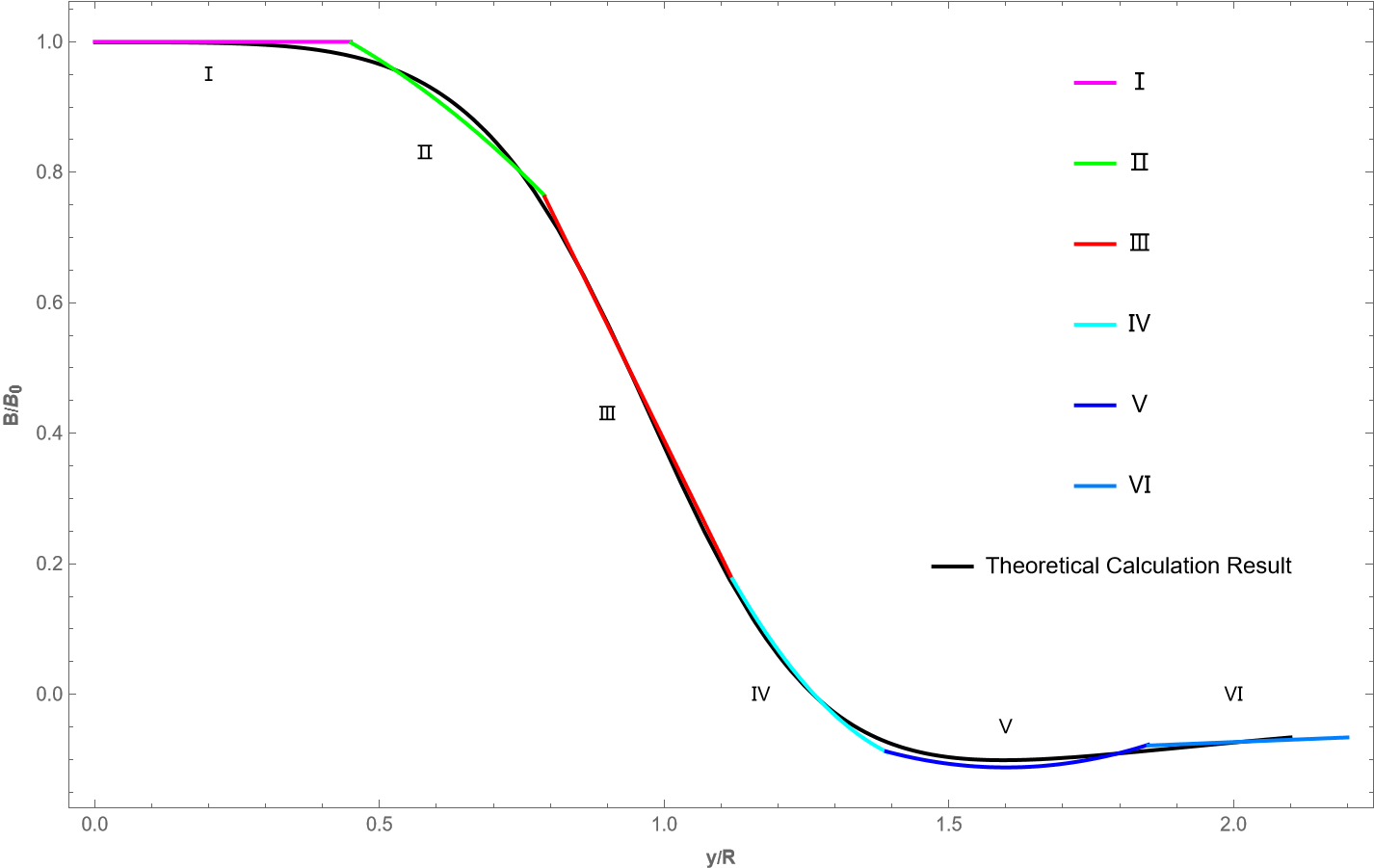}
    \caption{Results of piecewise function fitting for the theoretical simulation curve using elementary functions.}
    \label{fig:7}
\end{figure}

To simply the theoretical calculation of EMF on the metal rod under the magnetic field of Helmholtz coil, we employ the multiple simple functions to perform piecewise fitting for the theoretical black curve in figure~\ref{fig:7}. In principle, more piecewise segments yield a much smoother fitted curve and better agreement with the true profile. For practical convenience, six segments are used in this work. The fitted magnetic field distribution is given as:
\begin{equation}
\frac{B(r)}{B_0} =
\begin{cases}
1 \quad (\text{I}), & 0 \le r/R < R_{41} \\[4pt]
-0.5556\,r^{2} + 1.1111 \quad (\text{II}), & R_{41} \le r/R < R_{42} \\[6pt]
-1.7778\,r + 2.1667 \quad (\text{III}), & R_{42} \le r/R < R_{43} \\[6pt]
2\,(r-1.5)^{2} - 0.1122 \quad (\text{IV}), & R_{43} \le r/R < R_{44} \\[6pt]
0.5556\,(r-1.6)^{2} - 0.1122 \quad (\text{V}), & R_{44} \le r/R < R_{45} \\[6pt]
0.0356\,r - 0.1444 \quad (\text{VI}), & r/R \ge R_{45}
\end{cases}
\label{eq:7}
\end{equation}
where $R_{41}=0.4472$, $R_{42}=0.7876$, $R_{43}=1.1159$, $R_{44}=1.3886$, $R_{45}=1.8451$. The fitting result is demonstrated in figure~\ref{fig:7}.

From figure~\ref{fig:7}, we can see that the curve obtained by connecting six simple functions end-to-end is already very close to the theoretical simulation result of the magnetic field of Helmholtz coil, and these six functions can all be used to semi-analytically calculate the EMF on the metal rod. Of course, using more functions for fitting or using other simple functions for fitting may be more consistent with the theoretical curve, but this would make the EMF calculation very complicated. We will find that the results derived from these six functions are already in very good agreement with the experimental data, and the results of different function fittings do not differ very much. We will give the concrete calculation process below.

\section{Calculation of the induced EMF on the metal rod corresponding to the four magnetic field models}
\label{Sec4}

\subsection{Calculation of the induced EMF on the metal rod  under the step function model}

\begin{figure}
    \centering
    \hspace{2.5cm}  
    \includegraphics[width=0.7\linewidth]{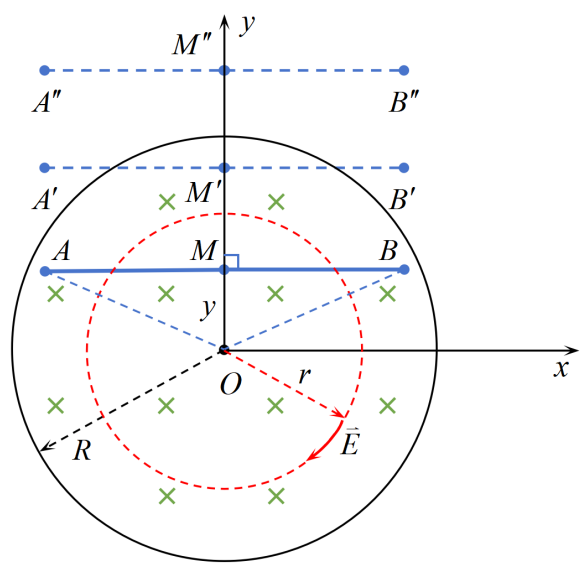}
    \caption{Thin metal rod $AB$ at different positions (along the $y$ direction) within the cylindrical magnetic field.}
    \label{fig:8}
\end{figure}

Under the step function model, the magnetic field of the Helmholtz coil is confined entirely within the coil volume, with uniform magnitude everywhere and orientation parallel to the axial direction. Its cross-section perpendicular to the $z$ axis is illustrated in figure~\ref{fig:8}. A thin metal rod is aligned along the $x$ axis and has a total length of $2l$. The center of the metal rod lies on the $y$ axis throughout the experiment. The two endpoints of the metal rod are given by $A(-l,y,0)$ and $B(l,y,0)$, as depicted in figure~\ref{fig:8}.
Owing to the axial symmetry of the magnetic field, the induced electric field generated in the surrounding space by a time-varying magnetic field also exhibits axial symmetry, as illustrated by the dashed circle of radius $r=\sqrt{x^{2}+y^{2}}$ in figure~\ref{fig:8}. On this circle, the induced electric field has identical magnitude at all points and points tangentially to the circle. The magnitude of the induced electric field produced by the time-dependent magnetic field at an arbitrary spatial position can be derived from Faraday's law of electromagnetic induction, we have
\begin{equation}
\oint_{l} \vec{E} \cdot d\vec{l} = -\frac{d\phi_{m}}{dt}
\label{eq:8}
\end{equation}
where $\phi_{m}$ is magnetic flux.
The magnitudes of the induced electric field at various locations inside and outside the magnetic field region in figure~\ref{fig:8} can be obtained as:
\begin{equation}
r < R,\quad E_{1} = -\frac{r}{2} B_{0} \frac{dB_{t}(t)}{dt}
\label{eq:9}
\end{equation}
\begin{equation}
r \ge R,\quad E_{2} = -\frac{R^{2}}{2r} B_{0} \frac{dB_{t}(t)}{dt}
\label{eq:10}
\end{equation}
This is a well-known result presented in electromagnetism textbooks~\cite{Branislav2011}. Here, we explicitly express the time-dependent component of the magnetic field according to equation~\eqref{eq:1}. Using the induced electric field distribution for different regions of the cylindrical magnetic field, the EMF across the metal rod can be evaluated by integration:
\begin{equation}
\varepsilon_{AB} = \int_{A}^{B} \vec{E} \cdot d\vec{l}
\label{eq:11}
\end{equation}
Equation~\eqref{eq:11} yields the induced EMF on the thin metal rod at an arbitrary position.

As the thin metal rod moves radially outward away from the magnetic field region, three distinct configurations will be encountered successively: the rod lies entirely within the magnetic field; both ends of the rod lie outside the magnetic field whereas its middle segment remains inside; and the rod is completely outside the magnetic field region. These three cases will be discussed in detail in the following subsections.

\paragraph{Case 1: the thin metal rod lies entirely within the magnetic field region.}

This case corresponds to segment AB in figure~\ref{fig:8}, with $y \le \sqrt{R^{2}-l^{2}}$. By virtue of the symmetry of the thin metal rod, the induced EMF between terminals A and B is twice the value for segment MB. Using equations.~\eqref{eq:9} and \eqref{eq:11}, we obtain
\begin{equation}
\varepsilon^1_{AB} = 2\!\left(\int_{0}^{l} E_{1} \cos\theta \, dx\right) = B_{0} \frac{dB_{t}(t)}{dt} y l
\label{eq:12}
\end{equation}

Since radii OA and OB are perpendicular to the orientation of the induced electric field, the line integrals of the induced electric field along OA and OB vanish according to Eq.~\eqref{eq:11}. Hence these two radial segments contribute nothing to the total EMF. The EMF arising from the variation of magnetic flux within triangle OAB equals the EMF across the thin metal rod AB. One may therefore compute the magnetic flux directly and evaluate the induced EMF via equation~\eqref{eq:8}; this approach is referred to as the area method. For a spatially uniform magnetic field where the area $S$ is time-independent, we have
\begin{equation}
\varepsilon^1_{AB} = \frac{d(BS)}{dt} = \frac{dB}{dt} 2\left(\frac{1}{2}yl\right) = B_{0} \frac{dB_{t}(t)}{dt} yl
\label{eq:13}
\end{equation}
Equations~\eqref{eq:12} and \eqref{eq:13} yield identical results from the direct integration of the vortex electric field and from the area method. The area method remains valid for non-uniform magnetic field distributions, provided that radii OA and OB stay perpendicular to the electric field direction, though an integral is then required for magnetic flux evaluation. It is also observed that when the thin metal rod is fully immersed in the magnetic field region, the magnitude of the induced EMF increases linearly with the distance $y$.

\paragraph{Case 2: Both ends of the thin metal rod lie outside the magnetic field region.}

Here $\sqrt{R^{2}-l^{2}} < y \le R$, corresponding to segment $A'B'$ in figure~\ref{fig:8}. Adopting the same approach as for Case~1, the induced EMF across the metal rod is given by
\begin{equation}
\varepsilon^2_{AB} = B_{0} \frac{dB_{t}(t)}{dt} \left( y\sqrt{R^{2}-y^{2}} + R^{2}\left(\arctan\frac{l}{y} - \arctan\frac{\sqrt{R^{2}-y^{2}}}{y}\right) \right)
\label{eq:14}
\end{equation}
Equation~\eqref{eq:14} characterizes how the magnitude of the induced EMF varies as the thin metal rod moves outward away from the magnetic field region for this configuration. The variation is no longer a simple monotonic increase or decrease.

\paragraph{Case 3: The thin metal rod lies entirely outside the magnetic field region, i.e., $y > R$.}

This configuration is denoted by segment $A''B''$ in figure~\ref{fig:8}. Following the same procedure, we obtain
\begin{equation}
\varepsilon^3_{AB} = B_{0} \frac{dB_{t}(t)}{dt} R^{2} \arctan\frac{l}{y}
\label{eq:15}
\end{equation}
Equation~\eqref{eq:15} indicates that for the rod fully outside the magnetic field region, the induced EMF decreases as the perpendicular distance $y$ increases. For convenience, we put the above three formulas together as follows:
\begin{equation}
\varepsilon_{AB} =
\left\{
\begin{array}{@{}l @{\hspace{5cm}} l@{}}
\displaystyle B_{0} \frac{dB_{t}(t)}{dt} yl, 
& \text{if } 0 \le y < \sqrt{R^{2}-l^{2}}, \\[16pt]
\multicolumn{2}{@{}l@{}}{%
\displaystyle B_{0} \frac{dB_{t}(t)}{dt} \left( y\sqrt{R^{2}-y^{2}} + R^{2}\left(\arctan\frac{l}{y} - \arctan\frac{\sqrt{R^{2}-y^{2}}}{y}\right) \right), 
} \\[6pt]
& \text{if } \sqrt{R^{2}-l^{2}} \le y < R, \\[16pt]
\displaystyle B_{0} \frac{dB_{t}(t)}{dt} R^{2} \arctan\frac{l}{y}, 
& \text{if } y \geq R.
\end{array}
\right.
\label{eq:16}
\end{equation}
The EMF curve for the metal rod predicted by the step function model is plotted as the black curve in figure~\ref{fig:emf}. It can be observed that the trend from the step function model agrees qualitatively with the experimental data. Even though the magnetic field distribution is discontinuous within this model, the EMF remains continuous: the curve rises monotonically from zero, reaches a peak with increasing $y$, and then decreases monotonically.

We next analyse possible extrema of the induced EMF as the distance $y$ varies. From the area method argument presented above, the magnitude of the induced EMF on the thin metal rod is proportional to the magnetic flux through triangle OAB within the cylindrical magnetic field region for all three ranges of $y$ (for a uniform magnetic field, the magnetic flux is proportional to area of triangle OAB). Physically, an extremum in the EMF corresponds to an extremum in the area. When the thin metal rod passes through the centre $O$, i.e., $y=0$, and the area of triangle OAB vanishes; accordingly, the induced EMF across the metal rod equals zero. This result can also be recovered by taking the limit of equation~\eqref{eq:12}:
\begin{equation*}
\lim_{y \to 0} B_{0} \frac{dB_{t}(t)}{dt} yl = 0
\end{equation*}
When the thin metal rod is at infinity, that is $y=+\infty$, taking the limit of equation.~\eqref{eq:15}, we get:
\begin{equation*}
\lim_{y \to +\infty} R^{2} B_{0} \frac{dB_{t}(t)}{dt} \arctan\frac{l}{y} = 0
\end{equation*}
The induced EMF is also zero in this limit. For other values of $y$, the induced EMF is non-zero and takes a finite value. To sum up, a maximum magnitude of the induced EMF can occur at a certain position of $y$. One may ask: where is the position ?

Numerical evaluation of equation~\eqref{eq:16} yields the EMF peak for the metal rod in the step function model at $y/R=0.9692$, corresponding to a radial position $y=10.18$~cm (Case 2). This represents an outward offset of approximately $3.84\%$ compared with the experimental peak $y_{\mathrm{exp}}=9.8$~cm, i.e., $y/R=0.933$. Some discrepancy exists between the theoretical predictions of the step function model and experimental results, which can be analysed with reference to figure~\ref{fig:5}. Good agreement between the step function model and experimental data is found only in the central region $y<0.3R$. The magnetic field drops discontinuously from a constant value to zero at $r=R$, with no intermediate smooth transition region, which produces deviations between theoretical and measured curves for $y>0.3R$. As shown in the preceding model results, the reverse magnetic field along the $z$ axis outside the Helmholtz coil starts to make contributions for $y>1.26R$, whereas the step function model assumes zero magnetic field outside the coil; hence substantial deviation arises between theoretical and experimental curves in this region. Nevertheless, built upon the physical picture of an infinitely long closely wound solenoid, the step function model possesses distinct merits. It not only facilitates straightforward theoretical computation but also provides a qualitative interpretation for the experimental data of the Helmholtz coil, yielding reasonable physical explanations for both the curve trend and peak position.

\subsection{Results of the induced EMF on the metal rod under the linear decay model}

According to the expression of magnetic field for the linear decay model, equation~\eqref{eq:5}, using a method similar to the previous subsection, we can obtain the semi-analytical result for the EMF on the metal rod as follows:
\begin{equation}
\varepsilon_{AB} = 2B_{0}\frac{dB_{t}(t)}{dt}
\left\{
\begin{array}{l}
\text{If }0 \le y < R_{21}: \\[4pt]
\displaystyle \frac{1}{2}y\sqrt{R_{21}^{2}-y^{2}} + \frac{1}{2}R_{21}^{2}\left(\arctan\frac{l}{y} - \arctan\frac{\sqrt{R_{21}^{2}-y^{2}}}{y}\right) \\[8pt]
\displaystyle \quad + \int_{R_{21}}^{\sqrt{l^{2}+y^{2}}}(-1.89r + 2.268)\left(\arctan\frac{l}{y} - \arctan\frac{\sqrt{r^{2}-y^{2}}}{y}\right)r\,dr, \\[16pt]
\text{If } R_{21} \le y < \sqrt{R_{22}^{2}-l^{2}}: \\[4pt]
\displaystyle \frac{1}{2}R_{21}^{2}\arctan\frac{l}{y} + \int_{R_{21}}^{y}(-1.89r + 2.268)\left(\arctan\frac{l}{y}\right)r\,dr \\[8pt]
\displaystyle \quad + \int_{y}^{\sqrt{l^{2}+y^{2}}}(-1.89r + 2.268)\left(\arctan\frac{l}{y} - \arctan\frac{\sqrt{r^{2}-y^{2}}}{y}\right)r\,dr, \\[16pt]
\text{If } \sqrt{R_{22}^{2}-l^{2}} \le y < R_{22}: \\[4pt]
\displaystyle \frac{1}{2}R_{21}^{2}\arctan\frac{l}{y} + \int_{R_{21}}^{y}(-1.89r + 2.268)\left(\arctan\frac{l}{y}\right)r\,dr \\[8pt]
\displaystyle \quad + \int_{y}^{R_{22}}(-1.89r + 2.268)\left(\arctan\frac{l}{y} - \arctan\frac{\sqrt{r^{2}-y^{2}}}{y}\right)r\,dr, \\[16pt]
\text{If } y \ge R_{22}: \\[4pt]
\displaystyle \frac{1}{2}R_{21}^{2}\arctan\frac{l}{y} + \int_{R_{21}}^{R_{22}}(-1.89r + 2.268)\left(\arctan\frac{l}{y}\right)r\,dr
\end{array}
\right.
\label{eq:17}
\end{equation}

The curve corresponding to equation~\eqref{eq:17} is plotted as the blue curve in figure~\ref{fig:emf}. It can be observed that the linear decay model achieves good agreement with experimental data for $y/R<1.2$. Compared with the step function model, the linear decay model yields values much closer to experimental measurements. Likewise, for $y/R>1.2$ in the outer region of the Helmholtz coil, theoretical predictions deviate substantially from experimental values owing to contributions from the reverse magnetic field. Numerical computation based on equation~\eqref{eq:17} gives the EMF peak of the linear-decay model at $y/R=0.9481$, i.e., $y=9.96$~cm, corresponding to an outward offset of merely $1.59\%$ relative to the experimental peak. The above analysis demonstrates that the linear decay model offers obvious improvement over the step function model in reproducing the experimental data.

\subsection{Results of the induced EMF on the metal rod under the exponential decay model}

According to the magnetic field distribution function of the exponential decay model, equation~\eqref{eq:6}, we can obtain the theoretical result for the EMF on the metal rod as follows:
\begin{equation}
\varepsilon_{AB} = 2B_{0}\frac{dB_{t}(t)}{dt}
\left\{
\begin{array}{l}
\text{If } 0 \le y < \sqrt{R_{31}^{2}-l^{2}}: \\[4pt]
\displaystyle \frac{1}{2}yl, \\[16pt]
\text{If } \sqrt{R_{31}^{2}-l^{2}} \le y < R_{31}: \\[4pt]
\displaystyle \frac{1}{2}y\sqrt{R_{31}^{2}-y^{2}} + \frac{1}{2}R_{31}^{2}\left(\arctan\frac{l}{y} - \arctan\frac{\sqrt{R_{31}^{2}-y^{2}}}{y}\right) \\[8pt]
\displaystyle \quad + \int_{R_{31}}^{\sqrt{l^{2}+y^{2}}}\left(0.606e^{\frac{R-r}{R_{00}}} - 0.2\right)\left(\arctan\frac{l}{y} - \arctan\frac{\sqrt{r^{2}-y^{2}}}{y}\right)r\,dr, \\[16pt]
\text{If } y \ge R_{31}: \\[4pt]
\displaystyle \frac{1}{2}R_{31}^{2}\arctan\frac{l}{y} + \int_{R_{31}}^{y}\left(0.606e^{\frac{R-r}{R_{00}}} - 0.2\right)\arctan\frac{l}{y}\,r\,dr \\[8pt]
\displaystyle \quad + \int_{y}^{\sqrt{l^{2}+y^{2}}}\left(0.606e^{\frac{R-r}{R_{00}}} - 0.2\right)\left(\arctan\frac{l}{y} - \arctan\frac{\sqrt{r^{2}-y^{2}}}{y}\right)r\,dr
\end{array}
\right.
\label{eq:18}
\end{equation}
The theoretical curve corresponding to equation~\eqref{eq:18} is plotted as the green curve in figure~\ref{fig:emf}. By incorporating the reverse magnetic field within the exponential decay model, the overall trend of its theoretical curve outperforms those of the step function model and the linear decay model. This model yields a peak at $y/R = 0.9581$, corresponding to $y = 10.06$~cm, with an outward offset of $2.65\%$ relative to the experimental peak.

\subsection{Results of the induced EMF on the metal rod under the piecewise function fitting model}

According to the magnetic field distribution function of the piecewise function fitting model, equation~\eqref{eq:7}, simple functions can be adopted to compute the induced EMF on the metal rod for this model. Nevertheless, the resulting expressions are rather complicated. For brevity, these results are presented in the appendix \ref{appendix}. The corresponding theoretical curve is plotted as the red curve in figure~\ref{fig:emf}. It can be seen that this model achieves the best agreement with experimental results. The piecewise function fitting model predicts a peak at $y/R=0.9673$, i.e., $y=10.16$~cm, with an outward offset of $3.64\%$ relative to the experimental peak. Even so, a small central segment of the metal rod still lies within the radius of the Helmholtz coil ($R=10.5$~cm). Obtained by fitting with six piecewise segments, the red curve in Fig.~2 matches the experimental results well. The evolution of EMF over the full radial range shows the highest consistency with experimental data. It also provides alternative evidence that the deviations of the step function model at far outer positions of the Helmholtz coil originate from the effect of the reverse magnetic field outside the coil. In terms of overall fitting performance, this model can satisfactorily reproduce our measured experimental data.

\subsection{Comparison of the four Models: from ``step'' to ``continuous decay'' explanation}

From figure~\ref{fig:emf}, we can observe that all four models can quantitatively reproduce the variation trend of experimental data and provide reasonable predictions for the position of the EMF peak. Within the region $y \le 0.3R$ (the near-central magnetic field region), the predictions of $\varepsilon_{AB}$ from the four models nearly overlap and show good agreement with experimental measurements. This demonstrates that the ``uniform field approximation'' is valid for simplified analysis and calculation in this region.

Furthermore, noticeable deviation between the theoretical curve of the step function model and experimental values emerges for $y > 0.3R$. This originates from the oversimplification of the step function model, which assumes a uniform magnetic field across the full radius of the Helmholtz coil. In reality, the magnetic field of the Helmholtz coil undergoes pronounced continuous decay for $y > 0.5R$. Since the step function model fails to capture this actual decay behaviour, it yields EMF predictions lower than the measured values. Deviations are also present for $0.3R < y < 0.5R$. When the centre of the metal rod lies within this interval, the finite length of the rod means that the magnetic field contribution to the induced EMF is an integral effect. Part of the metal rod extends outside the experimentally measured uniform field zone near the Helmholtz coil axis, which gives rise to discrepancies between theoretical calculations and experimental results.

Regarding peak positions, the maxima predicted by the step function, linear decay, exponential decay and piecewise function fitting models are all in close agreement with experiments. Whether the magnetic field decays continuously or abruptly from its maximum magnitude exerts little influence on the location of the EMF peak.

In the outer region $y > 1.26R$, much larger deviations appear for both the step function and linear decay models, as they neglect the effect of the external reverse magnetic field. By contrast, the exponential decay and piecewise function fitting models incorporate the reverse magnetic field. Nevertheless, the exponential decay model exhibits obvious deviations for large distances, whereas the piecewise function fitting model yields results that are consistent with experimental data over the whole range within the margin of error.

\section{Conclusion}
\label{Sec5}

In this work, we measure the evolution of EMF on a metal rod placed at different radial positions within the magnetic field of a Helmholtz coil. Four radial magnetic field models (step function, linear decay, exponential decay, and piecewise function fitting) are adopted to compute the theoretical induced EMF values at various positions. It is found that all four models can all qualitatively reproduce the experimental data. The induced EMF on the metal rod rises monotonically from zero to a maximum and then decreases monotonically. This trend is independent of the choice of magnetic field model, while the piecewise function fitting model achieves good quantitative agreement with experimental measurements within the margin of error, because it applies more fixed model parameters in theoretical calculation, which make the magnetic field of this model much closer to the real case.

In the central region ($r < 0.5R$), the interior of the Helmholtz coil can be approximated as a uniform field zone. Predictions from the linear decay, exponential decay and piecewise function fitting models nearly overlap with experimental data. Such uniform field regions have wide applications in the frontiers such as sensor magnetic field calibration. Within the decay transition region ($0.5R < r < 1.26R$), the magnetic field drops from its maximum. The experimentally measured EMF on the metal rod reaches its maximum before the rod fully moves out of the coil radius, i.e., $y_{\max} < R$. All four models are capable of reproducing both the magnitude and location of this maximum, showing generally good consistency with experiments. For the near outer coil region ($1.26R < r \le 2R$), the effect of the reverse magnetic field in the radial outer zone of the Helmholtz coil must be taken into account. The exponential decay and piecewise function fitting models yield much better agreement with experimental results. For $r > 2R$, only the piecewise function fitting model can quantitatively interpret the experimental data..

\raggedbottom

\section{Data availability statement}
The data that support the findings of this study are available from the corresponding author upon reasonable request.

\section{Acknowledgements}
This work is supported by the National Natural Science Foundation of China (Grant No.12275081).

\section*{Author Declarations}
\textbf{Conflict of Interest}

The authors have no conflicts of interest to disclose.

\appendix  
\renewcommand{\theequation}{A\arabic{equation}}
\setcounter{equation}{0}

\vspace*{-30pt} 
\section*{Appendix: Theoretical results of induced EMF on the metal rod calculated in piecewise function fitting model}


\begin{equation*}
\small
\varepsilon_{AB}= 2B_0\frac{\mathrm{d}B_t(t)}{\mathrm{d}t}
\left\{
\begin{array}{@{}l@{}}
\text{If } 0\le y<\sqrt{R_{42}^{2}-l^{2}}: \\[6pt]
\displaystyle
\frac{1}{2} y\sqrt{R_{41}^{2}-y^{2}}+\frac{1}{2} R_{41}^{2}\left(\arctan\frac{l}{y}-\arctan\frac{\sqrt{R_{41}^{2}-y^{2}}}{y}\right) \\[8pt]
\displaystyle
+\int_{R_{41}}^{\sqrt{y^{2}+l^{2}}}
\left(-0.5556\,r^{2}+1.1111\right)
\left(\arctan\frac{l}{y}-\arctan\frac{\sqrt{r^{2}-y^{2}}}{y}\right)
r\,\mathrm{d}r, \\[3ex]
\end{array}
\right.
\end{equation*}

\begin{equation*}
\small
\varepsilon_{AB}= 2B_0\frac{\mathrm{d}B_t(t)}{\mathrm{d}t}
\left\{
\begin{array}{@{}l@{}}
\text{If } \sqrt{R_{42}^{2}-l^{2}}\le y<R_{41}: \\[6pt]
\displaystyle
\frac{1}{2} y\sqrt{R_{41}^{2}-y^{2}}+\frac{1}{2} R_{41}^{2}\left(\arctan\frac{l}{y}-\arctan\frac{\sqrt{R_{41}^{2}-y^{2}}}{y}\right) \\[8pt]
\displaystyle
+\int_{R_{41}}^{R_{42}}
\left(-0.5556\,r^{2}+1.1111\right)
\left(\arctan\frac{l}{y}-\arctan\frac{\sqrt{R_{42}^{2}-y^{2}}}{y}\right)
r\,\mathrm{d}r \\[8pt]
\displaystyle
+\int_{R_{41}}^{R_{42}}
\left(-0.5556\,r^{2}+1.1111\right)
\left(\arctan\frac{\sqrt{R_{42}^{2}-y^{2}}}{y}-\arctan\frac{\sqrt{r^{2}-y^{2}}}{y}\right)
r\,\mathrm{d}r \\[8pt]
\displaystyle
+\int_{R_{42}}^{\sqrt{y^{2}+l^{2}}}
\left(-1.7778\,r^{2}+2.1667\right)
\left(\arctan\frac{l}{y}-\arctan\frac{\sqrt{r^{2}-y^{2}}}{y}\right)
r\,\mathrm{d}r
\end{array}
\right.
\end{equation*}

\begin{equation*}
\small
\varepsilon_{AB}= 2B_0\frac{\mathrm{d}B_t(t)}{\mathrm{d}t}
\left\{
\begin{array}{@{}l@{}}
\text{If } R_{41}\le y<R_{42}: \\[6pt]
\displaystyle
\frac{1}{2} R_{41}^{2}\arctan\frac{l}{y}
+\int_{R_{41}}^{y}
\left(-0.5556\,r^{2}+1.1111\right)
\arctan\frac{\sqrt{R_{42}^{2}-y^{2}}}{y}\,r\,\mathrm{d}r \\[8pt]
\displaystyle
+\int_{y}^{R_{42}}
\left(-0.5556\,r^{2}+1.1111\right)
\left(\arctan\frac{\sqrt{R_{42}^{2}-y^{2}}}{y}-\arctan\frac{\sqrt{r^{2}-y^{2}}}{y}\right)
r\,\mathrm{d}r \\[8pt]
\displaystyle
+\int_{R_{41}}^{R_{42}}
\left(-0.5556\,r^{2}+1.1111\right)
\left(\arctan\frac{l}{y}-\arctan\frac{\sqrt{R_{42}^{2}-y^{2}}}{y}\right)
r\,\mathrm{d}r \\[8pt]
\displaystyle
+\int_{R_{42}}^{\sqrt{y^{2}+l^{2}}}
\left(-1.7778\,r^{2}+2.1667\right)
\left(\arctan\frac{l}{y}-\arctan\frac{\sqrt{r^{2}-y^{2}}}{y}\right)
r\,\mathrm{d}r, \\[3ex]
\end{array}
\right.
\end{equation*}

\begin{equation*}
\small
\varepsilon_{AB}= 2B_0\frac{\mathrm{d}B_t(t)}{\mathrm{d}t}
\left\{
\begin{array}{@{}l@{}}
\text{If } R_{42}\le y<\sqrt{R_{43}^{2}-l^{2}}: \\[6pt]
\displaystyle
\frac{1}{2} R_{41}^{2}\arctan\frac{l}{y}
+\int_{R_{41}}^{R_{42}}
\left(-0.5556\,r^{2}+1.1111\right)
\arctan\frac{l}{y}\,r\,\mathrm{d}r \\[8pt]
\displaystyle
+\int_{R_{42}}^{y}
\left(-1.7778\,r^{2}+2.1667\right)
\arctan\frac{l}{y}\,r\,\mathrm{d}r \\[8pt]
\displaystyle
+\int_{y}^{\sqrt{y^{2}+l^{2}}}
\left(-1.7778\,r^{2}+2.1667\right)
\left(\arctan\frac{l}{y}-\arctan\frac{\sqrt{r^{2}-y^{2}}}{y}\right)
r\,\mathrm{d}r
\end{array}
\right.
\end{equation*}

\begin{equation*}
\small
\varepsilon_{AB}= 2B_0\frac{\mathrm{d}B_t(t)}{\mathrm{d}t}
\left\{
\begin{array}{@{}l@{}}
\text{If } \sqrt{R_{43}^{2}-l^{2}}\le y<R_{43}: \\[6pt]
\displaystyle
\frac{1}{2} R_{41}^{2}\arctan\frac{l}{y}
+\int_{R_{41}}^{R_{42}}
\left(-0.5556\,r^{2}+1.1111\right)
\arctan\frac{l}{y}\,r\,\mathrm{d}r \\[8pt]
\displaystyle
+\int_{R_{42}}^{y}
\left(-1.7778\,r^{2}+2.1667\right)
\arctan\frac{\sqrt{R_{43}^{2}-y^{2}}}{y}\,r\,\mathrm{d}r \\[8pt]
\displaystyle
+\int_{y}^{R_{43}}
\left(-1.7778\,r^{2}+2.1667\right)
\left(\arctan\frac{\sqrt{R_{43}^{2}-y^{2}}}{y}-\arctan\frac{\sqrt{r^{2}-y^{2}}}{y}\right)
r\,\mathrm{d}r \\[8pt]
\displaystyle
+\int_{R_{42}}^{R_{43}}
\left(-1.7778\,r^{2}+2.1667\right)
\left(\arctan\frac{l}{y}-\arctan\frac{\sqrt{R_{43}^{2}-y^{2}}}{y}\right)
r\,\mathrm{d}r \\[8pt]
\displaystyle
+\int_{R_{43}}^{\sqrt{y^{2}+l^{2}}}
\left[2\,(r-1.5)^{2}-0.1122\right]
\left(\arctan\frac{l}{y}-\arctan\frac{\sqrt{r^{2}-y^{2}}}{y}\right)
r\,\mathrm{d}r, \\[3ex]
\end{array}
\right.
\end{equation*}

\begin{equation*}
\small
\varepsilon_{AB}= 2B_0\frac{\mathrm{d}B_t(t)}{\mathrm{d}t}
\left\{
\begin{array}{@{}l@{}}
\text{If } R_{43}\le y<\sqrt{R_{44}^{2}-l^{2}}: \\[6pt]
\displaystyle
\frac{1}{2} R_{41}^{2}\arctan\frac{l}{y}
+\int_{R_{41}}^{R_{42}}
\left(-0.5556\,r^{2}+1.1111\right)
\arctan\frac{l}{y}\,r\,\mathrm{d}r \\[8pt]
\displaystyle
+\int_{R_{42}}^{R_{43}}
\left(-1.7778\,r^{2}+2.1667\right)
\arctan\frac{l}{y}\,r\,\mathrm{d}r \\[8pt]
\displaystyle
+\int_{R_{43}}^{y}
\left[2\,(r-1.5)^{2}-0.1122\right]
\arctan\frac{l}{y}\,r\,\mathrm{d}r \\[8pt]
\displaystyle
+\int_{y}^{\sqrt{y^{2}+l^{2}}}
\left[2\,(r-1.5)^{2}-0.1122\right]
\left(\arctan\frac{l}{y}-\arctan\frac{\sqrt{r^{2}-y^{2}}}{y}\right)
r\,\mathrm{d}r
\end{array}
\right.
\end{equation*}

\begin{equation*}
\small
\varepsilon_{AB}= 2B_0\frac{\mathrm{d}B_t(t)}{\mathrm{d}t}
\left\{
\begin{array}{@{}l@{}}
\text{If } R_{44}\le y<\sqrt{R_{45}^{2}-l^{2}}: \\[6pt]
\displaystyle
\frac{1}{2} R_{41}^{2}\arctan\frac{l}{y}
+\int_{R_{41}}^{R_{42}}
\left(-0.5556\,r^{2}+1.1111\right)
\arctan\frac{l}{y}\,r\,\mathrm{d}r \\[8pt]
\displaystyle
+\int_{R_{42}}^{R_{43}}
\left(-1.7778\,r^{2}+2.1667\right)
\arctan\frac{l}{y}\,r\,\mathrm{d}r \\[8pt]
\displaystyle
+\int_{R_{43}}^{R_{44}}
\left[2\,(r-1.5)^{2}-0.1122\right]
\arctan\frac{l}{y}\,r\,\mathrm{d}r \\[8pt]
\displaystyle
+\int_{R_{44}}^{y}
\left[0.5556\,(r-1.6)^{2}-0.1122\right]
\arctan\frac{\sqrt{R_{45}^{2}-y^{2}}}{y}\,r\,\mathrm{d}r \\[8pt]
\displaystyle
+\int_{y}^{R_{45}}
\left[0.5556\,(r-1.6)^{2}-0.1122\right]
\left(\arctan\frac{\sqrt{R_{45}^{2}-y^{2}}}{y}-\arctan\frac{\sqrt{r^{2}-y^{2}}}{y}\right)
r\,\mathrm{d}r \\[8pt]
\displaystyle
+\int_{R_{44}}^{R_{45}}
\left[0.5556\,(r-1.6)^{2}-0.1122\right]
\left(\arctan\frac{l}{y}-\arctan\frac{\sqrt{R_{45}^{2}-y^{2}}}{y}\right)
r\,\mathrm{d}r \\[8pt]
\displaystyle
+\int_{R_{45}}^{\sqrt{y^{2}+l^{2}}}
\left(0.0356\,r-0.1444\right)
\left(\arctan\frac{l}{y}-\arctan\frac{\sqrt{r^{2}-y^{2}}}{y}\right)
r\,\mathrm{d}r, \\[3ex]
\end{array}
\right.
\end{equation*}

\begin{equation*}
\small
\varepsilon_{AB}= 2B_0\frac{\mathrm{d}B_t(t)}{\mathrm{d}t}
\left\{
\begin{array}{@{}l@{}}
\text{If } R_{45}\le y<\sqrt{R_{45}^{2}-l^{2}}: \\[6pt]
\displaystyle
\frac{1}{2} R_{41}^{2}\arctan\frac{l}{y}
+\int_{R_{41}}^{R_{42}}
\left(-0.5556\,r^{2}+1.1111\right)
\arctan\frac{l}{y}\,r\,\mathrm{d}r \\[8pt]
\displaystyle
+\int_{R_{42}}^{R_{43}}
\left(-1.7778\,r^{2}+2.1667\right)
\arctan\frac{l}{y}\,r\,\mathrm{d}r \\[8pt]
\displaystyle
+\int_{R_{43}}^{R_{44}}
\left[2\,(r-1.5)^{2}-0.1122\right]
\arctan\frac{l}{y}\,r\,\mathrm{d}r \\[8pt]
\displaystyle
+\int_{R_{44}}^{R_{45}}
\left[0.5556\,(r-1.6)^{2}-0.1122\right]
\arctan\frac{l}{y}\,r\,\mathrm{d}r \\[8pt]
\displaystyle
+\int_{R_{45}}^{y}
\left(0.0356\,r-0.1444\right)
\arctan\frac{l}{y}\,r\,\mathrm{d}r \\[8pt]
\displaystyle
+\int_{y}^{\sqrt{y^{2}+l^{2}}}
\left(0.0356\,r-0.1444\right)
\left(\arctan\frac{l}{y}-\arctan\frac{\sqrt{r^{2}-y^{2}}}{y}\right)
r\,\mathrm{d}r
\end{array}
\right.
\end{equation*}

\begin{equation*}
\small
\varepsilon_{AB}= 2B_0\frac{\mathrm{d}B_t(t)}{\mathrm{d}t}
\left\{
\begin{array}{@{}l@{}}
\text{If } \sqrt{R_{45}^{2}-l^{2}}\le y<R_{45}: \\[6pt]
\displaystyle
\frac{1}{2} R_{41}^{2}\arctan\frac{l}{y}
+\int_{R_{41}}^{R_{42}}
\left(-0.5556\,r^{2}+1.1111\right)
\arctan\frac{l}{y}\,r\,\mathrm{d}r \\[8pt]
\displaystyle
+\int_{R_{42}}^{R_{43}}
\left(-1.7778\,r^{2}+2.1667\right)
\arctan\frac{l}{y}\,r\,\mathrm{d}r \\[8pt]
\displaystyle
+\int_{R_{43}}^{R_{44}}
\left[2\,(r-1.5)^{2}-0.1122\right]
\arctan\frac{l}{y}\,r\,\mathrm{d}r \\[8pt]
\displaystyle
+\int_{R_{44}}^{y}
\left[0.5556\,(r-1.6)^{2}-0.1122\right]
\arctan\frac{\sqrt{R_{45}^{2}-y^{2}}}{y}\,r\,\mathrm{d}r \\[8pt]
\displaystyle
+\int_{y}^{R_{45}}
\left[0.5556\,(r-1.6)^{2}-0.1122\right]
\left(\arctan\frac{\sqrt{R_{45}^{2}-y^{2}}}{y}-\arctan\frac{\sqrt{r^{2}-y^{2}}}{y}\right)
r\,\mathrm{d}r \\[8pt]
\displaystyle
+\int_{R_{44}}^{R_{45}}
\left[0.5556\,(r-1.6)^{2}-0.1122\right]
\left(\arctan\frac{l}{y}-\arctan\frac{\sqrt{R_{45}^{2}-y^{2}}}{y}\right)
r\,\mathrm{d}r \\[8pt]
\displaystyle
+\int_{R_{45}}^{\sqrt{y^{2}+l^{2}}}
\left(0.0356\,r-0.1444\right)
\left(\arctan\frac{l}{y}-\arctan\frac{\sqrt{r^{2}-y^{2}}}{y}\right)
r\,\mathrm{d}r, \\[3ex]
\end{array}
\right.
\end{equation*}
\normalsize

\begin{equation*}
\small
\varepsilon_{AB}= 2B_0\frac{\mathrm{d}B_t(t)}{\mathrm{d}t}
\left\{
\begin{array}{@{}l@{}}
\text{If } y\ge R_{45}: \\[6pt]
\displaystyle
\frac{1}{2} R_{41}^{2}\arctan\frac{l}{y}
+\int_{R_{41}}^{R_{42}}
\left(-0.5556\,r^{2}+1.1111\right)
\arctan\frac{l}{y}\,r\,\mathrm{d}r \\[8pt]
\displaystyle
+\int_{R_{42}}^{R_{43}}
\left(-1.7778\,r^{2}+2.1667\right)
\arctan\frac{l}{y}\,r\,\mathrm{d}r \\[8pt]
\displaystyle
+\int_{R_{43}}^{R_{44}}
\left[2\,(r-1.5)^{2}-0.1122\right]
\arctan\frac{l}{y}\,r\,\mathrm{d}r \\[8pt]
\displaystyle
+\int_{R_{44}}^{R_{45}}
\left[0.5556\,(r-1.6)^{2}-0.1122\right]
\arctan\frac{l}{y}\,r\,\mathrm{d}r \\[8pt]
\displaystyle
+\int_{R_{45}}^{y}
\left(0.0356\,r-0.1444\right)
\arctan\frac{l}{y}\,r\,\mathrm{d}r \\[8pt]
\displaystyle
+\int_{y}^{\sqrt{y^{2}+l^{2}}}
\left(0.0356\,r-0.1444\right)
\left(\arctan\frac{l}{y}-\arctan\frac{\sqrt{r^{2}-y^{2}}}{y}\right)
r\,\mathrm{d}r
\end{array}
\right.
\end{equation*}
\normalsize

\FloatBarrier

\bibliographystyle{unsrturl}
\bibliography{study}

\end{document}